# In Praise and in Criticism of the Model of Continuous Spontaneous Localization of the Wave-Function

*by Sofia D. Wechsler*

## Abstract

Different attempts to solve the measurement problem of the quantum mechanics (QM) by denying the collapse principle, and replacing it with changes in the quantum formalism, failed because the changes in the formalism lead to contradictions with QM predictions. To the difference, Ghirardi, Rimini and Weber took the collapse as a real phenomenon, and proposed a calculus by which the wave-function should undergo a sudden localization. Later on, Ghirardi, Pearle and Rimini came with a change of this calculus into the CSL (continuous spontaneous localization) model of collapse. Both these proposals rely on the experimental fact that the reduction of the wave-function occurs when the microscopic system encounters a macroscopic object and involves a big amount of its particles. Both of them also change the quantum formalism by introducing in the Schrödinger equation additional terms with noisy behavior. However, these terms have practically no influence as long as the studied system contains only one or a few components. Only when the amount of components is very big, these terms become significant and lead to the reduction of the wave-function to one of its components.

The present work has two purposes: 1) proving that the collapse postulate is unavoidable; 2) applying the CSL model to the process in a detector and showing step by step the modification of the wave-function, until reduction.

As a side detail, it is argued here that the noise cannot originate in some classical field, contrary to the thought/hope of some physicists, because no classical field is tailored by the wave-functions of entanglements.

## Abbreviations

| | |
|---|---|
| APP | = amount of perturbed particles |
| CSL | = continuous spontaneous localization |
| GPR | = Ghirardi, Pearle, Rimini |
| GRW | = Ghirardi, Rimini, Weber |
| LHS | = left hand side |
| OPWF | = one particle wave-function |
| PPC | = pair-producing collision |
| QM | = quantum mechanics |
| RHS | = right hand side |
| SSE | = stochastic Schrödinger equation |

## 1. Introduction

In a profound analysis of tests of quantum systems, [1], J. von Neumann concluded that once a quantum system in the initial state $\psi$ is tested (non-destructively) and produces the result $\lambda_k$, the system remains in a state $\phi_k$ with the property that any subsequent measurement of the system for the same observable, would produce the same result, $\lambda_k$ (see for instance page 138 in [1] ). This conclusion was always confirmed by the experiment. G. Lüders refined von Neumann's work examining cases with degenerate eigenvalues [2]. But the question remains what happens with the other components $\phi_1, \ldots, \phi_{k-1}, \phi_{k+1}, \ldots, \phi_N$ of the initial wave-function. Do they disappear, or do they continue to exist?

The complementary question concerns tests done exclusively for the result $\lambda_k$, in which the detector remains silent. For instance, in the test of a one particle wave-function (OPWF) comprising $N$ wave-packets space-separated, a detector is placed only on the wave-packet $\phi_k$. If the detector does not click, was $\phi_k$ destroyed?

Some scientists believe that $\phi_1, \ldots, \phi_{k-1}, \phi_{k+1}, \ldots, \phi_N$ in the first case, and $\phi_k$ in the second case, disappear. But neither von Neumann, nor Lüders, brought a rigorous proof that such a disappearance really occurs. So, the reduction of the wave-function, or 'collapse', remained just as a postulate. Lüders motivated:

> "statements on the change of state due to measurement do not arise out of quantum theory itself through the inclusion of the measurement apparatus in the Schrödinger equation. Measurement, an act of cognizance, adds an element not already contained in the formulation of quantum theory."

Other physicists, displeased by the enigmatic collapse postulate, launched 'interpretations' of the quantum mechanics (QM). Trying to explain the process of measurement of quantum systems without this postulate, they introduced modifications in the standard quantum formalism, or even additional universes. The price of altering the formalism was a contradiction with the quantum predictions for one or another experiment. So happened with the most popular interpretations, e.g. the mechanics of de Broglie and Bohm [3, 4],[1] the full/empty waves hypothesis – see for example [6] for explanation of the concept[2] – the consistent histories [7],[3] the transactional interpretation [8].[4]
All these interpretations ignored the well known experimental fact that the reduction of the wave-function occurs in the presence of a macroscopic object and perturbs so many of its particles until its macroscopic state changes.

There is one proposal which, to the difference from the above interpretations, took the collapse postulate 'seriously', and suggested a bridge between the quantum and the classical formalism. Ghirardi, Rimini and Weber (GRW) [9] thought that the wave-function of a quantum system might undergo at random times a sudden *shrinking* to a small region (localization). They too proposed changes in the Schrödinger equation; however, the changes have negligible effect on the quantum system containing a small number of components, and great effect – localization – when very many components are involved. Thus, in fact, the quantum formalism would not be changed.

---

[1] The de Broglie-Bohm interpretation of QM is based on the assumption of particles floating inside the wave-function and following continuous trajectories. The existence of such continuous trajectories was disproved in [5] section 3.

[2] From the proof against continuous trajectories in the section 3 of [5] one infers that a 'full wave' cannot follow a continuous trajectory. On the other hand, neither can it jump from one region to another one, space-separated, because in this case a wave-packet could be an empty wave when meeting a detector, and become a full wave later and trigger a subsequent detector. That would contradict our experiments on quantum systems.

[3] The consistent histories theory is mainly due to R. Griffith. Regrettably, it disobeys the quantum formalism. As an example, in the "histories" (13.7) in [7] it appears that after passing through a beam-splitter, the wave-function is truncated. Such a truncation is not allowed by the quantum formalism, since the transformation of the wave-function by a beam-splitter is unitary.

[4] Inspired by the Wheeler–Feynman absorber theory, J. G. Crammer proposed the hypothesis that both the emitter of a quantum system and the detector, emit a forward-in-time wave and a backward-in-time wave. The detection is supposed to occur if "handshake" occurs between the forward-in-time wave of the emitter and the backward-in-time wave of the detector. However, to the difference from the Wheeler–Feynman absorber theory in which the two waves superpose, in the formalism of the transaction interpretation appears the arithmetical product of the two waves. This replacement is motivated by claiming that the product gives the Born rule. But that is at variance with the quantum formalism in which the Born rule involves the *inner* product (an integration over the variables) of two waves, not their simple *arithmetical* product.

A few years later, Ghirardi, Pearle, and Rimini (GPR) came with a refined version of the GRW proposal, the 'continuous spontaneous localization' (CSL) [10],[5] by which the localization occurs progressively in time, instead of suddenly.[6]

The modification in the Schrödinger equation consists in adding a stochastic noise and non-linear terms. The stochasticity of the noise mimics the stochasticity of the result of the measurement. The magnitude of these supplementary terms increases with the number of particles involved.

The issue that the collapse occurs in the presence of many particles was put by D. Bedingham in a definite way, [17],

> "Our experience in the use of quantum theory tells us that the state reduction postulate should not be applied to a microscopic system consisting of a few elementary particles until it interacts with a macroscopic object such as a measuring device."

**Remark 1**: This text uses frequently the word 'particle', and it also appears in citations. Unless otherwise specified, this word means a simple quantum system, of one or a few components.

R. Feynman also described the effect of the involvement of a big amount of particles – section 2.3 of [18],

> "The classical approximation, however, corresponds to the case that the dimensions, masses, times, etc., are so large that S is enormous in relation to $\hbar$ (=1.05 × 10$^{-27}$ erg·sec). Then the phase of the contribution S/$\hbar$ is some very, very large angle … small changes of path will, generally, make enormous changes in phase, … The total contribution will then add to zero; for if one path makes a positive contribution, another infinitesimally close (on a classical scale) makes an equal negative contribution. … But for the special path $\bar{x}(t)$, for which S is an extremum, a small change in the path produces, in the first order at least, no change in S. All the contributions from the paths in this region are nearly in phase, …, and do not cancel out"

So, when the number of components of a system increases so much that it becomes a macroscopic object, the wave-function of the total system is destroyed. Unfortunately, Feynman's explanation covers only the particular case of a OPWF consisting in a single wave-packet, not in a superposition of a couple of wave-packets space-separated.

The present work focuses on the CSL formalism. This formalism is not regarded as an explanation of the collapse process, because it is not yet known which field is this noisy field. The stochastic Schrödinger equation (SSE) is regarded as the best tool for investigating the collapse, and what this text does is to apply the CSL model for following what happens in a detector during the detector process. *None of the 'interpretations' of QM is able to do such a thing.*

Applied to the process in a detector, the model predicts the expected predictions: as long as a small number of particles from the detector are entrained, no localization occurs. However, as this number increases the localization appears. It is dictated by the noise. Some evolutions of the noise enhance very much the number of involved particles and the detector clicks. Other evolutions of the noise begin, at a certain time, to reduce the number of involved particles, and in the end the detector remains unperturbed and does not click; also, the wave-packet which did not trigger the detector, is erased.

---

[5] In fact, the model in [10] is a continuation and enhancement of proposals of N. Gisin, [11], and of Pearle [12].

[6] This model is described also in the section 3 of [13], sections 7 and 8 of [14], section II F in [15], and section II C of [16].

Another purpose of this work is to prove that the disappearance of a wave-packet that didn't trigger a detector placed on its path, is *demanded* by the very QM formalism. It is strange that von Neumann, Lüders, the supporters of the CSL model, didn't try to prove that. The present work tries to remove this lacuna.

By the end of writing this work I was notified of a recent article, [19], which also studies the detection process with the CSL model, though, with a different approach and a more complicated setup. To the difference from the present work, which shows step by step the wave-function reduction as the process in a detector unfolds, in [19] is examined the current in a completely classical device, the battery of the electric circuit. This component does not come in contact with the wave-function. A brief examination of the analysis in [19] is done in the subsection 5.5.

The mysterious noise in the SSE challenged the supporters of the CSL model to ask which field might stand behind it. N. Gisin pointed out in [20] that the interaction appearing in SSE between the quantum system and this noise is similar with the interaction of a quantum system with an environment. L. Diósi advanced the idea that it may be a universal gravitational noise [21], [22], [23], but Ghirardi et al. criticized that proposal [24].
Personally, I don't believe that the noisy field may be classical, despite the hope expressed by Bassi and Ghirardi in [25], that the noise would be ultimately proved to be classical, e.g. gravitational. Although the present work does not deal with entanglements, it can be said in general that a classical field cannot tailor its noise in each type of experiment according to the respective wave-function. Even within a single type of experiment, the noise won't be tailored in each trial and trial according to the setup that each experimenter chooses, all the more that each choice is at the free will of the chooser. Worse than that, the requirements on the noise are bound to become harder if the experimenters' labs are in relative movement.

The rest of the text is organized as follows: section 2 describes the evolution of the detection in a macroscopic detector, taking as an example a proportional counter with gas ionization. Section 3 proves a couple of theorems in support of the necessity of the reduction postulate. Section 4 develops the mathematical tools for the treatment of a stochastic process, starting from the general Itô equation. Then it applies the tools on a microscopic system, showing that, in agreement with the principles of the CSL model, such a system won't undergo localization. Section 5 treats the detection process in a gas proportional counter. It is proved that as the number of the electron-ion pairs produced by ionization increases, the localization occurs. A difficulty is pointed to, which remains to be examined in future works. Section 6 contains conclusions.

## 2. The macroscopic detector

Typically, a macroscopic detector contains a sensitive material. When some particle enters the detector, some physical property of the respective type of particle interacts with the sensitive material producing a perturbation which is amplified in different ways, and a detection is reported. Or, if the wave-function consists in a quantum superposition and a detector is placed only on one of the wave-packets, in part of the trials of the experiment the detector remains silent.
In this text the proportional counter of cylindrical symmetry is taken as example of detector, figure 1. The detailed description of this apparatus and its functioning can be found in [26 – 28].
A gas, usually of the noble type as He, Ar, Xe, or others, is placed between two electrodes of opposite electrical charges. The incident particle ionizes a couple of atoms in the gas generating so-called 'primary electrons and ions'.

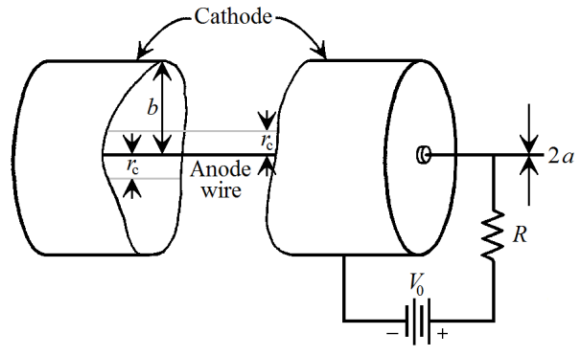

Figure 1. A proportional counter of cylindrical symmetry.
(See explanations in the text)

Both the electrons and the ions are accelerated by the electric field, though the electrons, having the mass at least $10^4$ times smaller than an ion, get a correspondingly higher acceleration. Thus, within the time necessary to the electron for reaching the anode, the ion practically doesn't move.

In the cylindrical counter the anode is a thin metallic wire passing through the center of the tube. Thus, the electric field increases as the distance from the anode decreases. At a certain distance $r_c$ from the central axis the field becomes so intense that the 'primary electrons' mentioned above have gained enough high kinetic energy for ionizing other atoms – figure 2a. Thus, the number of free electrons is doubled. The free electrons available after this first generation of so-called 'secondary ionizations', are quickly accelerated and undergo a second generation of pair-producing collisions (PPCs), doubling again the number of free electrons.

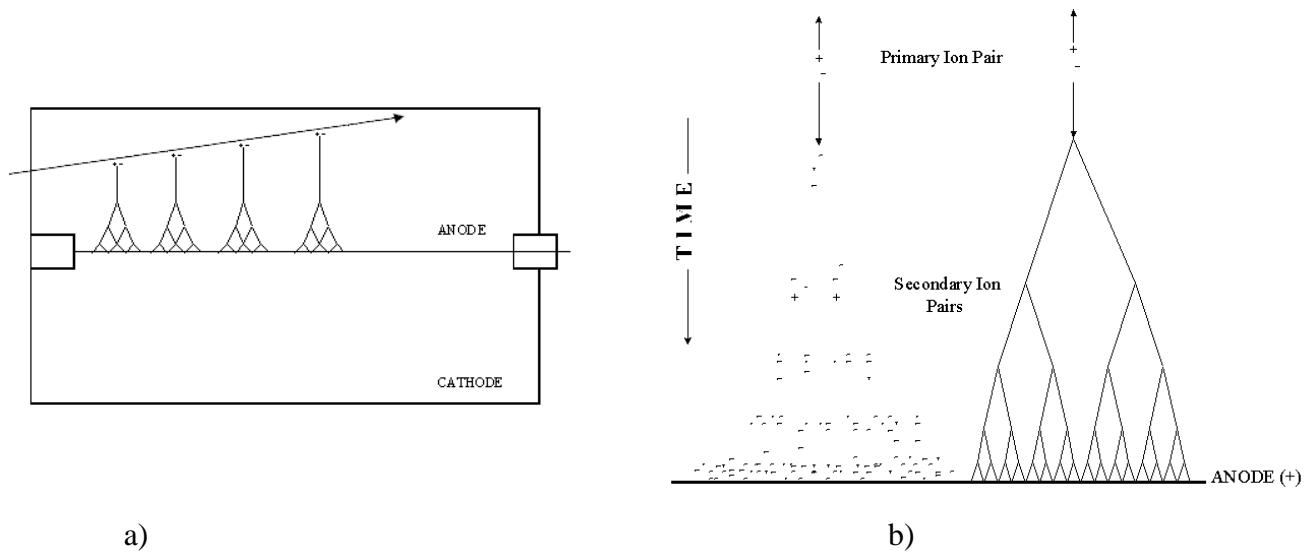

a)                                                                                  b)

Figure 2. Amplification in proportional detectors.

Both figures a and b are taken from the reference [27]. a) An energetic incident particle collides with the atoms in the counter ionizing some of them. The electrons set free by the direct interaction of the visiting particle with atoms, move toward the anode and in the avalanche region $r < r_c$ their energy increases enough for ionizing additional atoms, producing additional electron-ion pairs. The electrons are accelerated until another generation of ionizations occurs. b) The process continues in the same way, each generation of ionizations doubling the number of free electrons. After a couple of such generations, whole avalanches are created.

The process continues this way, each primary electron producing in short time an 'avalanche' – figure 2b – named in the literature 'Townsend avalanche', and the number of secondary electrons per primary electron is called multiplication factor. The region $r < r_c$ is called the 'avalanche region'.

After a certain amount of pairs are produced, the (almost) static cloud of ions in the avalanche region limits the intensity of the field, so, the production of additional pairs is stopped. The electrons produced so far are absorbed by the anode, entailing a short output signal with steep rise. The positive ions drift slowly toward the cathode producing an additional signal, much longer and rising more slowly.

Some atoms may absorb the energy of the hitting electrons, but instead of undergoing ionization they move to upper levels from which they de-excite with emission of energetic photons. Such a photon interacts further with an atom and sets free an electron by the photo-electric effect. The electron starts an additional avalanche, in the same way as a primary electron did. However, in a proportional counter such a process has negligible probability.

## 3. The collapse principle is unavoidable

The purpose of the present section is to prove that the principle of wave-function reduction is unavoidable, that, although the QM formalism cannot describe the collapse, denying the collapse leads to contradictions. We assume that the detectors are *ideal*. We will consider a OPWF with a couple of wave-packets, and for the beginning we will restrict our study to the simplest case: only one of the wave-packets illuminates a detector. One of two events will occur: the detector clicks, or, it 'responds' by silence.

## 3.1. Measurements in which the detector remains silent

How could a detector remain non-impressed by a wave-packet?
Many scenarios can be proposed. It is impossible to deal with all the suggestions the imagination can advance, all the more that some suggestions may contradict in different ways the laws of physics. There are three options which seem to me worth of consideration:

1) The perturbation produced by the wave-packet in the detector is too small for changing the macroscopic state of the material in the detector.

2) The wave-packet is retro-flected by the detector wall and escapes through the input window.

3) The wave-packet doesn't see the detector. (How so, I don't know.)

The following theorems rule out these options.

***Theorem 1:***
*If a wave-packet meets a detector and does not trigger it, no particle in the detector remains perturbed.*

***Proof:***
From a down-conversion pair of photons, the idler photon is sent to a detector Q, and the signal photon is sent to a detector S – figure 3. The signal wave-packet is split by the 50-50% beam-splitter BS, into two copies. one reflected, $|a\rangle$, and one transmitted, $|b\rangle$. Both copies travel to a rotating mirror M, initially in horizontal position. The paths from the nonlinear crystal which produces the pair (not shown in the figure) to the detector Q and to the mirror M are tuned so that the idler reaches and triggers Q just after the wave-packet $|a\rangle$ was reflected by M. Upon the click of Q, M is rotated to vertical position. Thus, $|b\rangle$, which is retarded comparatively to $|a\rangle$ by the mirrors m, does not meet M, and continues its travel on the same track as. At this step the wave-function of the signal photon becomes, considering all the reflections at mirrors,

$$|\phi\rangle = -(|a\rangle + \mathrm{i}|b\rangle)/\sqrt{2} . \tag{1}$$

In continuation, the wave-packets $|a\rangle$ and $|b\rangle$ follow their common track toward the detector S. The wave-packet $|a\rangle$ is the first one to meet the detector. At the interaction with a first particle (atom/molecule) from the sensitive material in the detector, an entanglement is generated

$$|\phi\rangle|A_1^{\mathrm{u}}\rangle \rightarrow -(|a^{\mathrm{p}}\rangle|A_1^{\mathrm{p}}\rangle + \mathrm{i}|b^{\mathrm{u}}\rangle|A_1^{\mathrm{u}}\rangle)/\sqrt{2} , \tag{2}$$

where the upper-script 'p' means 'perturbed', and 'u' means 'unperturbed'.
The first term between the round parentheses on the RHS indicates that both the incoming particle and the one from the detector are perturbed. Further, they meet and perturb additional particles as explained in the previous section. As long as the number of perturbed particles is small, their total system continues to be described by the quantum formalism,

$$|\Phi\rangle = -(|a^{\mathrm{p}}\rangle|A_1^{\mathrm{p}}\rangle|A_2^{\mathrm{p}}\rangle \ldots + \mathrm{i}|b^{\mathrm{u}}\rangle|A_1^{\mathrm{u}}\rangle|A_2^{\mathrm{u}}\rangle \ldots)/\sqrt{2} , \tag{3}$$

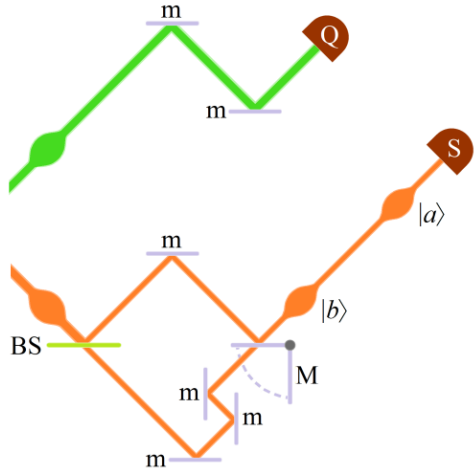

Figure 3. A which-way experiment.
The colors in the figure are only for eye-guiding. BS is a fair beam-splitter, m are fixed mirrors, M is a rotatable mirror, S and Q are ideal detectors. From a down-conversion pair, the idler photon (green) travels toward the detector Q, and the signal photon (orange) lands on a fair beam-splitter BS, where its wave-packet is split into a reflected part $|a\rangle$, and a transmitted part $|b\rangle$. The mirror M is initially in horizontal position, thus, $|a\rangle$ is reflected and directed toward the detector S. Immediately after that, the idler photon meets the detector Q which clicks. Upon this event, M is rotated to vertical. So, the retarded wave-packet $|b\rangle$ doesn't encounter the mirror M and travels towards the detector S, along the same track as $|a\rangle$.

Assume that meanwhile the wave-packet $|b\rangle$ also reaches the detector S. It starts a new chain of perturbations. For simplicity, we consider the two sets of perturbed particles, disjoint.

Let $t$ be a time at which $|a\rangle$ engaged $N$ perturbed particles, and $|b\rangle$, $M$ perturbed particles, with both $N$ and $M$ still small enough for allowing a quantum description. The equation (3) evolved into

$$|\Theta\rangle = -\big(|a^{\mathrm{p}}\rangle|A_1^{\mathrm{p}}\rangle\ldots|A_N^{\mathrm{p}}\rangle|A'^{\mathrm{u}}_1\rangle\ldots|A'^{\mathrm{u}}_M\rangle + \mathrm{i}|b^{\mathrm{p}}\rangle|A_1^{\mathrm{u}}\rangle\ldots|A_N^{\mathrm{u}}\rangle|A'^{\mathrm{p}}_1\rangle\ldots|A'^{\mathrm{p}}_M\rangle\big)/\sqrt{2}\,. \qquad (4)$$

The detector particles correlated with $|b\rangle$ were marked with prime for distinguishing them from those correlated with $|a\rangle$.

Now, let's see what the RHS of (4) tells us.

One can see that if the particles $A'_1\ldots A'_M$ are perturbed, then $A_1\ldots A_N$ should not have been perturbed, therefore they could not begin an avalanche. So, the wave-packet $|a\rangle$ left the detector silent. Setting in particular $N=1$, one infers that if $A'_1\ldots A'_M$ are perturbed, the wave-packet $|a\rangle$ that didn't trigger the detector, did not perturb even one single particle of the detector..

Symmetrically, if the particles $A_1\ldots A_N$ were perturbed, then $A'_1\ldots A'_M$ are forced to remain unperturbed, so they cannot start avalanche. Therefore, the wave-packet $|b\rangle$ does not trigger the detector. Setting in particular $M=1$, one infers that if $A_1\ldots A_N$ were perturbed, $|b\rangle$ that didn't impress the detector did not perturb even one particle of the detector.

That confirms the theorem and rules out the option (1).

### Theorem 2:
*A wave-packet that did not impress in an ideal, absorbing detector, was destroyed.*

### Proof:

As the ideal detector has perfectly reflecting walls, the escape can be only by retro-flection on the walls and exit through the input window.

Consider again a pair of down-conversion photons, in which the idler is sent to a detector Q, and the signal to a beam-splitter BS – figure 4. At BS, the signal wave-packet is split into a reflected copy $|a\rangle$, and a transmitted copy $|b\rangle$. The latter flies in continuation through the vicinity of a rotating mirror M, initially parallel to the path of $|b\rangle$. Then, $|b\rangle$ is reflected by a fixed mirror m and flies toward a detector S.

If $|b\rangle$ does not impress the detector, the theorem 1 says that it does not perturb even one single particle in the detector.

The length of the path from the nonlinear crystal which produces the pair (not shown in the figure) to the detector Q, is equal to that to the mirror m. Upon the click of Q, the mirror M is rotated counter-clockwise by $90°$. If $|b\rangle$ is retro-flected inside the detector toward the input window, as says the assumption 2 in the beginning of this subsection, it returns to the mirror m where it is reflected toward the mirror M which is now perpendicular to the path of $|b\rangle$. So, $|b\rangle$ is reflected back to m and from there it returns to the detector S.

As far as is known to me, the experiment proposed here was not performed. However, there are other experiments, which showed that the repetition of a test of an observable on the same quantum system, produces the same result as the first test, [29]. This is also the prediction of the QM itself.

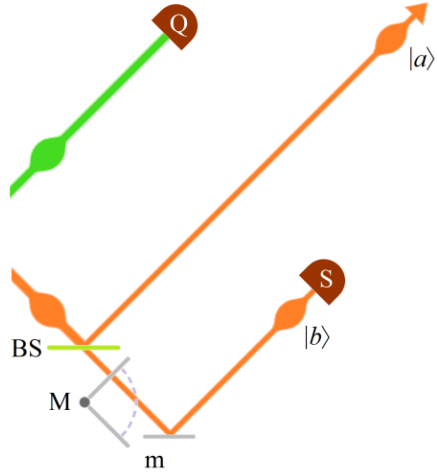

Figure 4. Test of the possibility of retro-flection of a wave-packet by an ideal detector.
The colors in the figure are only for eye-guiding. BS is a fair beam-splitter, m is a fixed mirror, M is a rotatable mirror, S and Q are ideal, absorbing detectors. From a down-conversion pair, the idler photon (green) travels toward the detector Q, and the signal photon (orange) to BS. The mirror M is initially parallel to the path of the transmitted beam $|b\rangle$, therefore the beam does not meet the mirror. Then, $|b\rangle$ travels on and is reflected by the fixed mirror m. At this time, the idler photon meets the detector Q which clicks. Upon the click of Q, M is rotated to a position perpendicular to the path between BS and m. The wave-packet $|a\rangle$ reflected by BS, travels freely for a by far longer time than needed to $|b\rangle$ to meet the detector S. If, as said in the text, the wave-packet $|b\rangle$ succeeds to exit the detector and returns to the mirror m, this time it will be reflected by the mirror M and sent back to m, and from here to S.

So, if $|b\rangle$ is indeed retro-flected and returns to the detector, neither this time would the detector click. The hypothesis that the silence in S is due to retro-flection of $|b\rangle$ entails that $|b\rangle$ is again sent to M, and it returns to the detector. In this way, the round trip of $|b\rangle$ would be repeated endlessly.

However, this is impossible. An ideal, absorbing detector cannot be traversed endlessly, and the wave-packet never interact even with one particle inside.
In particular, even if the wave-packet does not exit the detector, but just traverses the detector being reflected from wall to wall, it can't traverse endlessly the detector and never interact even with one particle in detector.

Thus, the theorem is proved, ruling out the option (2).

### Theorem 3:
*It's impossible that a wave-packet which met an ideal absorbing detector, without triggering it, didn't feel the detector presence and passed on, unchanged.*

Before proceeding to the proof let's notice that it is impossible to assume that a wave-packet that didn't trigger an ideal absorbing detector, passes on *distorted*, because the distortion should be the effect of the encounter with particles in the detector. However, the theorem 1 says that these particles remain unperturbed.

### Proof:
The proof will be done by disproving the opposite, that the wave-packet exited the detector unchanged.
The experimental arrangement bears some similarity to the Elitzur-Vaidman interaction-free measurement [30], figure 5. From a down-conversion pair, the signal photon lands on the input beam-splitter BS of a Mach-

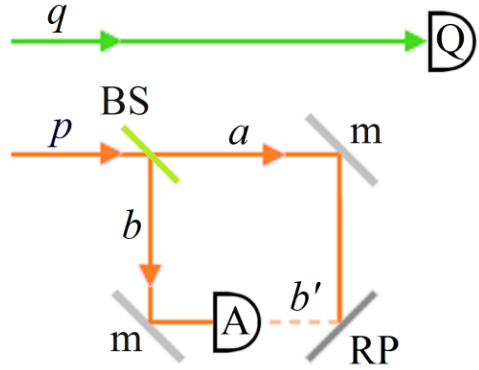

Figure 5. An experiment of Elitzur-Vaidman type.

The colors are only for eye-guiding. From an idler-signal pair of down-conversion photons the idler (green) is sent to a detector Q for heralding the presence of the signal photon (orange) in the apparatus. The signal is sent to a 50%-50% beam-splitter BS. m are fixed mirrors, A and Q are ideal, absorbing detectors. RP is a recording (e.g. photographic) plate. The significance of the path $b'$ – dashed, pale orange – is explained in the text.

Zehnder interferometer. The idler photon is sent to a detector Q for heralding the presence of the signal photon. The signal photon exits BS in the state

$$|\psi\rangle = \left( |a\rangle + i|b\rangle \right) / \sqrt{2} \,. \tag{5}$$

After being reflected by the mirrors m, $|a\rangle$ and $|b\rangle$ cross one another on a recording plate RP (for instance, a photographic plate). The internal arms of the interferometer are of equal length. Therefore, if the detector A were absent, $|a\rangle$ and $|b\rangle$ would have left in each trial of the experiment a spot on the plate, and in the end of the experiment an interference pattern would have appeared on RP.

However, since A is in place, the situation is different: the experiments say that in the trials in which both detectors Q and A click, the RP is not impressed; it is impressed only in the trials in which Q clicks and A doesn't, and no interference image appears on the plate.

Let's though see what would be the implications if in the trials in which Q clicks and A doesn't, $|b\rangle$ just didn't feel the detector presence and passed on, unchanged. Let's name the wave-packet beyond A, $|b'\rangle$. The theorem 1 says that if A doesn't click, no particle in A is perturbed. The wave-function describing the result is

$$|\Phi\rangle = i\left( |a\rangle|A_1^u\rangle|A_2^u\rangle\ldots + i|b'\rangle|A_1^u\rangle|A_2^u\rangle\ldots \right) / \sqrt{2} = i|A_1^u\rangle|A_2^u\rangle\ldots\left( |a\rangle + i|b'\rangle \right) / \sqrt{2} \,. \tag{6}$$

So, the signal photon would be described by

$$|\phi\rangle = \left( |a\rangle + i|b'\rangle \right) / \sqrt{2} \,. \tag{7}$$

This wave-function would produce on the RP an interference pattern.

However, the QM doesn't predict such an effect.

These theorems leave a strange image: a wave-packet that enters a detector and does not produce a click, does not disturb even one particle of the detector. If the detector is absorbing and ideal, it is not possible that the wave-packet returns repeatedly to the detector, or remains inside and endlessly traverses it, though does not perturb even one particle inside. Neither does the wave-packet continue its path past the detector.

The only explanation that seems to remain is that the wave-packet is destroyed by the detector.

## 3.2. Measurements in which the detector clicks – a non-decidable problem

So far we dealt with the fate of wave-packets which don't impress detectors. However, there exists also the complementary problem: if the wave-function has, say, a couple of wave-packets, e.g. $|\psi_1\rangle$, $|\psi_2\rangle$, $|\psi_3\rangle$, and $|\psi_1\rangle$ triggers a detector $D_1$, what happens with $|\psi_2\rangle$ and $|\psi_3\rangle$? The experiment shows that if we place detectors on them, $D_2$, respectively $D_3$, these detectors remain silent. Therefore, according to the conclusion of the previous subsection, $|\psi_2\rangle$ and $|\psi_3\rangle$ disappear. The question is, when do they disappear?

Let's assume that the detectors are not permanently in the setup, that we introduce them when we wish, e.g. first $D_1$ and after that, $D_2$ and $D_3$. Did $|\psi_2\rangle$ and $|\psi_3\rangle$ disappear when $D_1$ clicked, or later, when they met $D_2$ and $D_3$?

Although this work does not deal with relativity, it can be said in general that if the relativistic intervals between the click in $D_1$ and the impingements in $D_2$, respectively $D_3$ are space-like, one can find frames of coordinates by which the order of the events is opposite.

Let $\mathcal{F}_1$ be a frame by which $D_1$ is introduced first, and $\mathcal{F}_2$ a frame by which $D_2$ and $D_3$ are introduced first. If, judging according to the time axis of $\mathcal{F}_1$, we assume that when $D_1$ clicks, $|\psi_2\rangle$ and $|\psi_3\rangle$ disappear, then, according to $\mathcal{F}_2$ no wave-packets impinged on $D_2$ and $D_3$, i.e. $|\psi_2\rangle$ and $|\psi_3\rangle$ did not exist at all. However, we produced a wave-function with three wave-packets, and their existence does not depend on the frame by which we do calculi.

Now, let's see what can say the non-relativistic quantum mechanics.

Two types of tests can be done for detecting a wave-packet: i) a which-way test, or, ii) interference with another wave-packet. The experiments show that in which-way tests, if a wave-packet caused a click in a detector, the other detectors remain silent. But these tests give no answer to the question whether the click in one detector erased the other wave-packets (collapse), or, the silent detectors destroy by themselves the impinging wave-packets.

What remains is to see whether interference experiments can answer.

***Theorem 4***. *If a wave-packet of a OPWF makes a detector click, the other wave-packets don't show any interference effect.*

***Proof:***

Figure 6 illustrates a reduced version of the Sciarrino-type experiment [31]. On two 50%-50% beam-splitters, $BS_A$ and $BS_B$, land identical photons from a degenerate-down-conversion pair. From each photon, the reflected part $|a\rangle$ ($|b\rangle$) is sent to an ideal, absorbing detector, A (B), and the transmitted part $|a'\rangle$ ($|b'\rangle$) is sent to a detector array, DA, consisting in miniature absorbing detectors.[7] The device DA records for each absorbed photon the coordinates in the plane of the plate and the recording time.

---

[7] Such a net of detectors was used in the experiments described in [32].

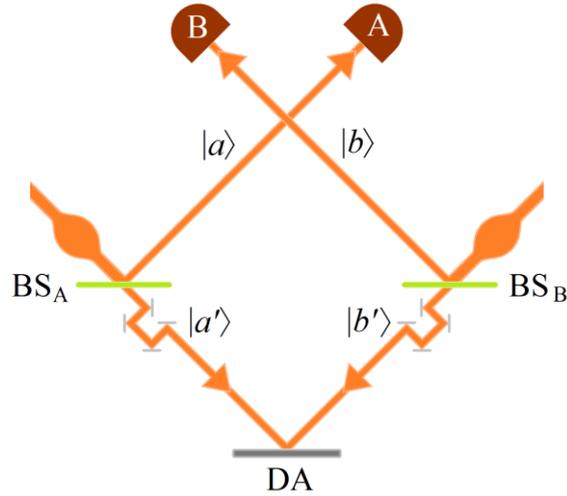

Figure 6. A Sciarrino-type experiment.

The colors are only for eye-guiding. BS$_A$ and BS$_B$ are 50%-50% beam-splitters. They are illuminated by identical photons from a degenerate-down-conversion pair. A and B are ideal, absorbing detectors. DA is a 2D recording array of miniature, ideal, absorbing detectors, which record the time and the x-y coordinates of the absorbed photon. The small, light-grey lines are mirrors which introduce a delay such that the optical distance from the beam-splitter to the detector A (B) is equal to that to DA.

The joint wave-function of the two photons is

$$|\varPhi\rangle = \frac{1}{2}\,(\,\mathrm{i}|a\rangle + |a'\rangle)(\mathrm{i}|b\rangle + |b'\rangle) = \frac{1}{\sqrt{2}}\left(\mathrm{i}\frac{|a\rangle|b'\rangle + |b\rangle|a'\rangle}{\sqrt{2}} - \frac{|a\rangle|b\rangle - |a'\rangle|b'\rangle}{\sqrt{2}}\right). \tag{8}$$

We will discard the trials ending with a joint detection in A and B, or, with a double detection in the array DA. So, we will work only with the (truncated and renormalized) state

$$|\psi\rangle = (|a\rangle|b'\rangle + |b\rangle|a'\rangle) / \sqrt{2}\,. \tag{9}$$

Let's consider the cases when A clicks and B remains silent.

As A is an ideal absorbing detector, this click testifies that the wave-packet $|a\rangle$ was absorbed, therefore, what exits A is vacuum. Also, since B remained silent, according to the former subsection the wave-packet $|b\rangle$ should have been destroyed, leaving beyond this detector also vacuum. Introducing these inferences in (9) one gets

$$|\phi\rangle = \big(|0\rangle|b'\rangle + |0\rangle|a'\rangle\big) / \sqrt{2} = |0\rangle\big(|b'\rangle + |a'\rangle\big) / \sqrt{2}\,. \tag{10}$$

It seems that in these cases the wave-packets $|a'\rangle$ and $|b'\rangle$ should produce on DA an interference pattern, confirming that although the wave-packet $|a\rangle$ was detected, $|a'\rangle$ survived.

Absolutely the same tableau should appear from trials ending with a detection in B and no detection in A.

However, the wave-functions (10) is not correct. It ignores the involvement of the particles in the detector A (B). As long as the *quantum description* can still be used, (10) should be replaced by

$$|\Phi\rangle_{\mathrm{A}} = |0\rangle\big(|b'\rangle|A_1^{\mathrm{p}}\rangle|A_2^{\mathrm{p}}\rangle\ldots + |a'\rangle|A_1^{\mathrm{u}}\rangle|A_2^{\mathrm{u}}\rangle\ldots\big)/\sqrt{2}\,, \tag{11}$$

respectively,

$$|\Phi\rangle_{\mathrm{B}} = |0\rangle\big(|b'\rangle|A_1^{\mathrm{u}}\rangle|A_2^{\mathrm{u}}\rangle\ldots + |a'\rangle|A_1^{\mathrm{p}}\rangle|A_2^{\mathrm{p}}\rangle\ldots\big)/\sqrt{2}\,, \tag{12}$$

where the superscript 'u' stands for unperturbed, and 'p' for perturbed. (Of course, when the number of particles perturbed in the clicking detector becomes very big, the coherent superposition in the RHS of (11) and (12) is broken, and only the term with the perturbed particles remains.)

Bottom line, no interference tableau can appear.

Thus, the theorem 4 is proved.

## 4. Stochastic evolution of a microscopic system

In the section 2.1 of [10], GPR developed an equation for a stochastically evolving system described by a state-function $|\psi\rangle$. They started from a very general equation of this type, the Itô equation,

$$\mathrm{d}|\psi\rangle = \Big(\hat{C}\,\mathrm{d}t + \sum_n \hat{A}_n\,\mathrm{d}B_n\Big)|\psi\rangle\,. \tag{13}$$

In this equation $\hat{C}$ and $\{\hat{A}_n\}$ are operators, and $\{B_n\}$ is a set of real, Wiener processes.[8] Since in this text we are going to work with a single operator $\hat{A}$ we rewrite (13) as,

$$\mathrm{d}|\psi\rangle = \big(\hat{C}\,\mathrm{d}t + \hat{A}\,\mathrm{d}B\big)|\psi\rangle\,. \tag{14}$$

A Wiener process refers to a stochastic quantity $B$ which may take at any time $t$ from the beginning of a given trial of the experiment, an arbitrary value. If $\mathrm{d}t$ is a small interval of time, after each such $\mathrm{d}t$ the quantity $B$ may change value. The differential element $\mathrm{d}B(t) = B(t+\mathrm{d}t) - B(t)$ obeys certain constraints. If $B$ is a white noise the constraints are:

$$\overline{\mathrm{d}B(t)} = 0\,,\ \ \mathrm{i})\qquad \overline{[\mathrm{d}B(t)]^2} = \gamma\,\mathrm{d}t\,,\ \ \mathrm{ii}) \tag{15}$$

where the average is taken over all the trails, at the time $t$ measured since the beginning of each trial. $\gamma$ is a parameter expressing the intensity of the noise.

Replacing the non-Hermitian part of the operator $\hat{C}$ with $-\mathrm{i}\hat{H}$, where $\hat{H}$ is the Hamiltonian, the equations (13) and (14) would look like a Schrödinger equation, though with additional terms besides the

---

[8] A Wiener process refers to a parameter $B(t)$ which varies continuously in time, however, the value of the increment $\mathrm{d}B(t)$ jumps from time to time, so that the derivative of $B(t)$ has points of discontinuity.

Hamiltonian. The presence of additional terms yields a different solution that that of the Schrödinger equation, and this solution is non-normalized. For normalizing the solution

$$|\phi\rangle_t = |\psi\rangle_t / [_t\langle\psi|\psi\rangle_t]^{1/2}, \tag{16}$$

GPR did the appropriate changes in (13). Next, applying the Itô formalism [33], they obtained a non-linear SSE, which for a single Hermitian operator $\hat{A}$, reads

$$d|\phi\rangle_t = \left\{ -i\hat{H}\,dt - \frac{1}{2}\gamma\,[\hat{A} - R(t)]^2\,dt + [\hat{A} - R(t)]\,dB(t) \right\}|\phi\rangle_t, \ \ \text{i)} \qquad R(t) = {}_t\langle\phi|\hat{A}|\phi\rangle_t. \ \ \text{ii)}^9 \tag{17}$$

**Remark 2:** The equation (17) reduces to the Schrödinger equation if the last two terms in (17i) bring a negligible contribution to the solution, in comparison with $i\hat{H}\,dt$. It will be seen in the end of this section and in the next section that as long as a system consists in a few microscopic components the last two terms have indeed a negligible effect.

**Remark 3:** It was proved in [13] that a nonlinear equation would allow faster than light communication. However, the stochastic character of the noise impedes such a possibility.

Further, GPR tried to prove – section 2.2. of [10] – that a system evolving by the non-linear SSE obtained by them – ends up in an eigenfunction of the set $\{\hat{A}_n\}$. By analogy, our equation (17) with a single operator, $\hat{A}$, should have as solution one of the eigenstates of $\hat{A}$. The purpose of this section and of the next one is to check in detail this inference.

GPR decided to ignore in the influence of $\hat{H}$ during the process of localization, motivating:

"Since we are interested here in discussing the physical effects of the new terms, we disregard for the moment the Schrödinger part of the dynamical equation."

This decision is not always correct. *If $\hat{H}$ contains interaction terms responsible for the evolution of the system toward localization, $\hat{H}$ cannot be just ignored.* This will be the case in this text. Fortunately, it will be shown that $\hat{H}$ cancels out in the calculi.

For proving that the CSL model predicts localization, GPR followed a procedure that is exposed below with a single operator $\hat{A}$, and without ignoring the Hamiltonian *a priori*.

Expanding the wave-function $|\phi\rangle$ according to the eigenstates $|a_j\rangle$ of the operator $\hat{A}$ one gets

$$|\phi\rangle = \sum_j a_j|\phi\rangle. \tag{18}$$

Denoting by $\hat{P}_j$ be the projection operator on the eigenstate $|a_j\rangle$,

---

9 Detailed explanations of these calculi are given in the section 7 of [14].

$$|\phi\rangle = \sum_j \hat{P}_j|\phi\rangle \, . \tag{19}$$

Obviously,

$$\hat{P}_j|\phi\rangle = \boldsymbol{a}_j(t)|a_j\rangle \, , \tag{20}$$

$$\langle\phi|\hat{P}_j|\phi\rangle = |\boldsymbol{a}_j(t)|^2 \, . \tag{21}$$

From (19) one immediately obtains

$$\hat{A}|\phi\rangle = \sum_j a_j \, \hat{P}_j|\phi\rangle \, , \tag{22}$$

$$\hat{A}^2|\phi\rangle = \sum_j a_j^2 \, \hat{P}_j|\phi\rangle \, , \tag{23}$$

$$R(t) = \sum_j \langle\phi|a_j \, \hat{P}_j|\phi\rangle = \sum_j a_j \, \langle\phi|\hat{P}_j|\phi\rangle \, . \tag{24}$$

It is more useful for our purpose to work with the probabilities than with the amplitudes. Introducing (19) in (17), using the equalities (22–24), and projecting on the eigenstate $|a_m\rangle$,

$$\mathrm{d}\hat{P}_m|\phi\rangle = \left\{ -\mathrm{i}E_m \, \mathrm{d}t - \frac{1}{2}\gamma[a_m - R(t)]^2 \, \mathrm{d}t + [a_m - R(t)]\,\mathrm{d}B(t) \right\} \hat{P}_m|\phi\rangle \, . \tag{25}$$

The following identity results from the rules of the Itô stochastic calculus – see also section 7.3 in [14],

$$\mathrm{d}\langle\phi|\hat{P}_m|\phi\rangle = \left[\mathrm{d}\langle\phi|\hat{P}_m\right]|\hat{P}_m|\phi\rangle + \langle\phi|\hat{P}_m\left[\mathrm{d}\,\hat{P}_m|\phi\rangle\right] + \left[\mathrm{d}\langle\phi|\hat{P}_m\right]\left[\mathrm{d}\,\hat{P}_m|\phi\rangle\right] \, . \tag{26}$$

We will calculate the components of this equation using (19), (22 − 25) and the properties (15) of d$B$.

$$\mathrm{d}\langle\phi|\hat{P}_m|\phi\rangle = \mathrm{d}[|\boldsymbol{a}_m(t)|^2] \, ,$$
$$\left[\mathrm{d}\langle\phi|\hat{P}_m\right]|\hat{P}_m|\phi\rangle + \langle\phi|\hat{P}_m\left[\mathrm{d}\,\hat{P}_m|\phi\rangle\right] = \left\{ -\gamma[a_m - R(t)]^2 \, \mathrm{d}t + 2[a_m - R(t)]\,\mathrm{d}B(t) \right\}|\boldsymbol{a}_m(t)|^2 \, , \tag{27}$$
$$\left[\mathrm{d}\langle\phi|\hat{P}_m\right]\left[\mathrm{d}\,\hat{P}_m|\phi\rangle\right] \approx [a_m - R(t)]^2 \, |\boldsymbol{a}_m(t)|^2 \, \gamma \, \mathrm{d}t \, .$$

In the calculus of $\left[\mathrm{d}\langle\phi|\hat{P}_m\right]\left[\mathrm{d}\,\hat{P}_m|\phi\rangle\right]$ we ignored terms with $(\mathrm{d}t)^2$ and the terms with d$B$d$t$ vis-à-vis the terms with d$B$ and the terms with d$t$ since d$t$ is usually extremely small.
Let's notice that the terms with the Hamiltonian cancelled out one another.
Introducing the RHSs of (27) in (26) there results

$$\mathrm{d}[|\boldsymbol{a}_m(t)|^2] = 2|\boldsymbol{a}_m(t)|^2 \, [a_m - R(t)]\,\mathrm{d}B(t) \, . \tag{28}$$

One may be eluded by the form of this equation and think that dividing both sides by $|\boldsymbol{a}_m(t)|^2$, it is possible to integrate and get,

$$|\boldsymbol{a}_m(t)|^2 = |\boldsymbol{a}_m(t_0)|^2 \exp\left\{2\int\limits_{t_0}^{t}\left[a_m - R(\tau)\right] \mathrm{d}B(\tau)\right\}. \tag{29}$$

However, (21) and (24) show that $R(t)$ also depends on the set $\{|\boldsymbol{a}_j(t)|^2\}$. Besides that, $\mathrm{d}B(t)$ is not a known function, $\mathrm{d}B$ changes at each time arbitrarily, within the constraints (15).

Then, the solution of (28) should be calculated iteratively, each iteration corresponding to a $\mathrm{d}t$, and for each iteration one should pick at random a value for $\mathrm{d}B$, within the constraints (15). One can start from $t_0$, a time before the noise began to act on the quantum system, so all $\{|\boldsymbol{a}_j(t_0)|^2\}$ are known from the initial wave-function. The first iteration would calculate the set of differentials $\{\mathrm{d}[|\boldsymbol{a}_j|^2]\}$ using (28). The second iteration will update $\{|\boldsymbol{a}_j|^2\}$ using the set $\{\mathrm{d}[|\boldsymbol{a}_j|^2]\}$ and will calculate a new set $\{\mathrm{d}[|\boldsymbol{a}_j|^2]\}$. In the next iteration $|\boldsymbol{a}_j|^2$ will be again updated, and so on. A specific example is given in the subsection 5.2.

Though, the formula (29) is useful for getting some bounds of $\{|\boldsymbol{a}_j(t)|^2\}$ during its evolution. Let's check for instance, if the intensity $|\boldsymbol{a}_m(t)|^2$ may vanish. For such an end, the value of the exponent in the RHS of (29) should be $-\infty$. In practice, if this exponent accumulates up to a time $t_1$ a value, say, –13, it would render $|\boldsymbol{a}_m(t_1)|^2$ quite small, of the order of $|\boldsymbol{a}_m(t_0)|^2 \times 2.3 \times 10^{-6}$.

Admitting that the eigenvalues $\{a_j\}$ are dimensionless and of the order of unity, e.g. the spin projection of a spin 1 boson (with $\hbar$ set equal to 1), $R$ would be of the same order of magnitude.

According to (15ii), $|\mathrm{d}B|$ fluctuates around $\sqrt{\gamma\,\mathrm{d}t}$. In the section 5 we will give a numerical example which *imposes* $\mathrm{d}t$ to be about $5 \times 10^{-13}\,\mathrm{s}$, and $\gamma = 800\mathrm{s}^{-1}$ which is by very many orders of magnitude greater than $\gamma$ values provided by experiments on the CSL model [34]. So, a simple choice for $\mathrm{d}B$ would be $\pm 2 \times 10^{-5}$.

Despite the constraint (15i) which allows $\mathrm{d}B$ to change sign even after each $\mathrm{d}t$, let's choose for $\mathrm{d}B$ the sign opposite to $[a_m - R(\tau)]$ in each interval $\mathrm{d}t$.

With these data, if the exponent in the RHS of (29) accumulates until $t_1$ the value –13, the integral in (29) is the sum of $6.5/(2 \times 10^{-5}) = 3.25 \times 10^5$ pieces. As each such piece lasts $5 \times 10^{-13}\,\mathrm{s}$, the decrease of $|\boldsymbol{a}_m|^2$ would take $3.25 \times 10^5 \times 5 \times 10^{-13}\,\mathrm{s} = 162.5\mathrm{ns}$. In some detectors, the detection time is much shorter, so that the exponent doesn't have the time to accumulate a sufficient value. For comparison, as we will see in the subsection 5.3, in a proportional counter it takes $0.1\mathrm{ns}$ until the output signal begins to rise, i.e. cca. $1.6 \times 10^3$ times less. And that, besides the fact that the constraint (15i) does not allow $\mathrm{d}B$ to be always of sign opposite to $[a_m - R]$.

The conclusion is not that the CSL model is wrong, to the contrary. Since the calculus above was done for a simple system, e.g. one quantum particle, it was expected according to the very principle of the model that no localization can occur. As the model says, the localization appears in a multi-component system. A totally different situation will be described in the next section.

## 5. The CSL model and the detection process

In this section the CSL model is going to be tested on the process occurring inside a detector when a quantum particle enters and more and more particles from the detector are perturbed. As said in section 2, we do our rationale on a detector working in the proportional regime. It will be supposed that the initial wave-function is a OPWF of the form

$$/\Psi\rangle_{t_0} = \boldsymbol{a}_1(t_0)/\Psi_1\rangle + \boldsymbol{a}_2(t_0)/\Psi_2\rangle + \boldsymbol{a}_3(t_0)/\Psi_3\rangle, \quad \text{i)} \qquad \sum_{k=1}^{3} /\boldsymbol{a}_k(t)/^2 = 1, \quad \text{ii)} \tag{30}$$

where $/\Psi_k\rangle$ indicates the presence of a quantum particle on the path $k$. It will also be supposed that an ideal detector is placed on the wave-packet $/\Psi_1\rangle$; no detectors will be placed on the other wave-packets. It is expected that the model make the following predictions:

**a)** If the detector clicks, the intensity of the wave-packet $/\Psi_1\rangle$ has increased to 1. The number of electron-ion pairs in the detector becomes equal to the number of pairs that would have been produced if the initial wave-function possessed only one wave-packet, $/\Psi_1\rangle$, (so that the detection by the ideal detector would have been sure).

**b)** If the detector remains silent, $/\boldsymbol{a}_1(t)/^2$ decreases to zero according to the theorem 1 which requires that the number of electron-ion pairs in the detector, become null.

### 5.1. The test of a many-component system – the process inside a detector

With the impingement of $/\Psi_1\rangle$ on the detector, the OPWF changes into an entanglement. As described in section 2, the visitor wave-packet may cause a couple of primary ionizations, and then, each primary electron is dragged by the electric field to the multiplication region. Thus, the entanglement takes the form

$$/\Phi\rangle_t = \boldsymbol{a}_1(t)\sum_{j=1}^{\infty} c_j(t)/j\rangle + \boldsymbol{a}_2(t)/\Psi_2\rangle/0\rangle + \boldsymbol{a}_3(t)/\Psi_3\rangle/0\rangle, \quad \text{i)} \qquad \sum_{j=0}^{\infty} /c_j(t)/^2 = 1, \quad \text{ii)} \tag{31}$$

where $j$ is the number of electron-ions pairs in the detector. The upper limit of $j$ is the number of atoms in the detector, a number known only as an order of magnitude. But this number is huge in comparison with the number of electron-ion pairs that may appear, for which reason it was set to $\infty$.

GPR developed an equation for the evolution of a set of identical particles – section 3.1 in [10],

$$\mathrm{d}/\psi\rangle = \left\{ -\mathrm{i}\hat{H}\,\mathrm{d}t - \frac{1}{2}\gamma \int \mathrm{d}\mathbf{r}\,\hat{N}^2(\mathbf{r})\,\mathrm{d}t + \int \mathrm{d}\mathbf{r}\,\hat{N}(\mathbf{r})\,\mathrm{d}B(\mathbf{r},t) \right\}/\psi\rangle, \tag{32}$$

where the operator $\hat{N}(\mathbf{r})$ was attached to the density of number of particles at the position $\mathbf{r}$, and $\langle\hat{N}(\mathbf{r})\rangle_t$ is the average number of particles at the position $\mathbf{r}$ and time $t$. Details on how $\hat{N}(\mathbf{r})$ can look like, on its eigenfunctions and eigenvalues, can be found in [14] section 8.1 and [10] section 3.1.

Normalizing the wave-function (32) as we did in the former section – relation (16) – one gets

$$\mathrm{d}/\phi\rangle_t = \left\{-\mathrm{i}\hat{H}\,\mathrm{d}t - \frac{1}{2}\gamma\int\mathrm{d}\mathbf{r}\left[\hat{N}(\mathbf{r})-\langle\hat{N}(\mathbf{r})\rangle_t\right]^2\mathrm{d}t + \int\mathrm{d}\mathbf{r}\left[\hat{N}(\mathbf{r})-\langle\hat{N}(\mathbf{r})\rangle_t\right]\mathrm{d}B(\mathbf{r},t)\right\}/\phi\rangle_t. \qquad (33)$$

GPR obtained the equation (32) for a fixed number of particles. However, in a detector the situation is different; the visitor particle perturbs more and more particles from the detector as the time passes. In the case of a proportional counter, inside the avalanche region the number of electron-ion pairs increases exponentially in time as explained in section 2. That would require that the Hamiltonian comprise creation and annihilation operators, e.g.

$$\hat{G} = \hat{a}_\mathrm{e}^\dagger(\mathbf{r},t)\,\hat{a}_\mathrm{ion}^\dagger(\mathbf{r},t)\,\hat{a}_\mathrm{atom}(\mathbf{r},t) + \mathrm{c.c.}, \qquad (34)$$

and the increase in time of the number of PPCs would make $\hat{H}$ more and more complicated. Fortunately, in our calculi the Hamiltonian will be eliminated, as it was eliminated in the calculi in section 4.

Another problem is that in part of the collisions are not produced electron-ion pairs but excited atoms on high levels. Then, the atoms de-excite through emission of high energy photons, which set free new electrons by the photo-electric effect. These electrons ionize atoms, starting secondary avalanches, i.e. avalanches which do not originate in primary electrons.

For avoiding these complications we shall make a couple of assumptions, part of them adopted also in [26] chapter 6, section III A.

\*) Since the probability of secondary avalanches in the proportional counter is small, we assume that no free electron is lost, each free electron undergoes a PPC.

"The only multiplication process is through electron collisions (any photoelectric effects are neglected), that no electrons are lost to negative ion formation, and that space-charge effects are negligible". [26]

\*\*) In the present study there is no benefit from assuming that d$B$ is position dependent inside the detector. This fact will appear more convincing in the subsection 5.2. Therefore, inside the detector d$B$ will be considered as varying only in time.

The space-distribution of the electrons and ions inside the detector is also going to be irrelevant in our analysis. Given also the assumption (\*\*), we can work with a simpler equation

$$\mathrm{d}/\Phi\rangle_t = \left\{-\mathrm{i}\hat{H}\,\mathrm{d}t - \frac{1}{2}\gamma\left(\hat{N}-\langle\hat{N}\rangle_t\right)^2\mathrm{d}t + \left(\hat{N}-\langle\hat{N}\rangle_t\right)\mathrm{d}B(t)\right\}/\Phi\rangle_t, \quad \mathrm{i}) \qquad \langle\hat{N}\rangle_t = {}_t\langle\Phi/\hat{N}/\Phi\rangle_t. \quad \mathrm{ii}) \quad (35)$$

where the operator $\hat{N}$ is associated with the total number of electron-ion pairs in the detector, $\hat{N}/j\rangle = j/j\rangle$. Applying it to the wave-function (31i) and calculating $\langle\hat{N}\rangle_t$ according to (35ii)

$$\langle \hat{N} \rangle_t = /\boldsymbol{a}_1(t)/^2 \, Y(t) \,, \quad \text{i)} \qquad Y(t) = \sum_{j=1}^{\infty} j / c_j(t)/^2 \,. \quad \text{ii)} \tag{36}$$

**Remark 4:** the formulas (36) may eventually confuse the reader, so, it is better to stress the difference between $\langle \hat{N} \rangle_t$ and $Y(t)$. As one can infer from (36i), $Y(t)$ is number of electron-ion pairs that *would be present in the detector* if $/\boldsymbol{a}_1(t)/^2$ were equal to 1, i.e. if the wave-function consisted in only one wave-packet, $/\Psi_1\rangle$. In this case the ideal detector would have reported the incident particle, with certainty. However, the wave-function contains three wave-packets, not one. The wave-function (31) tells us that the $Y(t)$ pairs could be found in the detector, with the probability $/\boldsymbol{a}_1(t)/^2$, and with a probability $1 - /\boldsymbol{a}_1(t)/^2$ no pairs would be found. $\langle \hat{N} \rangle_t$ takes in consideration this situation.

In continuation, for finding out whether the model can predict (**a**) or (**b**), we try to get an equation for the evolution in time of $/\boldsymbol{a}_1/^2$. We begin by projecting the equation (35i) onto the Fock state of $m$ pairs,

$$\mathrm{d}\,\hat{P}_m/\Phi\rangle_t = \hat{P}_m \mathrm{d}/\Phi\rangle_t = -\left\{ \mathrm{i}E_m \, \mathrm{d}t + \frac{1}{2}\gamma \big[ \langle \hat{N} \rangle_t - m \big]^2 \mathrm{d}t + \big[ \langle \hat{N} \rangle_t - m \big] \mathrm{d}B(t) \right\} \hat{P}_m/\Phi\rangle_t \,, \tag{37}$$

and use that in the identity (26). The result is similar with (28) with the only difference that – see (31) – instead of (20) and (21) one has

$$\hat{P}_m/\Phi\rangle_t = \boldsymbol{a}_1(t)\, c_m(t)/m\rangle \,, \quad \text{i)} \qquad \langle \Phi \,|\, \hat{P}_m/\Phi\rangle_t = /\boldsymbol{a}_1(t)/^2/c_m(t)/^2 \,. \quad \text{ii)} \tag{38}$$

Doing this replacement in (28),

$$\big[ \mathrm{d}/\boldsymbol{a}_1(t)/^2 \big]/c_m(t)/^2 + /\boldsymbol{a}_1(t)/^2 \, \mathrm{d}/c_m(t)/^2 = 2 \big[ m - \langle \hat{N} \rangle_t \big]/c_m(t)/^2 /\boldsymbol{a}_1(t)/^2 \, \mathrm{d}B(t) \,. \tag{39}$$

For getting rid of $\mathrm{d}/c_m(t)/^2$ we do summation over $m$, use (31ii) and (36ii), then, replace $\langle \hat{N} \rangle_t$ by (36i),

$$\mathrm{d}/\boldsymbol{a}_1(t)/^2 = 2 \big[ 1 - /\boldsymbol{a}_1(t)/^2 \big] \, Y(t) /\boldsymbol{a}_1(t)/^2 \, \mathrm{d}B(t) \,. \tag{40}$$

## 5.2. General implications

Comparing the equation (40) with (28), it's obvious that the presence of the number of particles $Y(t)$ in the former – see remark 4 – makes $/\boldsymbol{a}_1(t)/^2$ vary in much bigger steps than in the case of one single particle. According to the explanations in section 2 and the assumption (*), $Y$ increases exponentially in time,

$$Y_k = n_0 \, 2^k \,, \tag{41}$$

where $n_0$ is the number of the primary electrons and $k$ is the number of the last PPC generation.

If the physical conditions constraint

$$0 \leq /\boldsymbol{a}_1(t)/^2 \leq 1 \,, \tag{42}$$

is obeyed during the detection process, the equation (40) shows that $/\boldsymbol{a}_1/^2$ would increase if $dB > 0$, and decrease if $dB < 0$, since $Y$ is positive. For sufficiently high values of $Y$, $dB > 0$ would augment $/\boldsymbol{a}_1/^2$ toward $1$ – expectation (**a**), while $dB < 0$ would lower $/\boldsymbol{a}_1/^2$ toward $0$ – expectation (**b**).
A specific example is given in the next subsection.

It would be appealing to divide on both sides of (40) by $\left[1 - /\boldsymbol{a}_1(t)/^2\right]/\boldsymbol{a}_1(t)/^2$ and integrate the equation, if $dB$ were a known function. But it isn't. What one can do is to solve (40) iteratively, by assuming some sequence $dB_1, dB_2, \ldots$

However, the validity of (40) is limited: if from some iteration results $d/\boldsymbol{a}_1/^2 < 0$ with $\left|d/\boldsymbol{a}_1/^2\right| > /\boldsymbol{a}_1/^2$, in the next iteration one will get $/\boldsymbol{a}_1/^2$ negative, which is non-physical. The following limitation emerges straightforwardly from (40)

$$/2\,Y(t)\left[1 - /\boldsymbol{a}_1(t)/^2\right]dB(t)/ < 1 \,. \tag{43}$$

If in some iteration there results $d/\boldsymbol{a}_1/^2 > \left(1 - /\boldsymbol{a}_1/^2\right)$, in the next iteration $/\boldsymbol{a}_1/^2$ would exceed 1, which is non-physical too. Obviously, if $d/\boldsymbol{a}_1/^2 > \left(1 - /\boldsymbol{a}_1/^2\right)$, $d/\boldsymbol{a}_1/^2$ is positive. There emerges immediately from (40)

$$0 < 2\,Y(t)/\boldsymbol{a}_1(t)/^2\,dB(t) < 1 \,. \tag{44}$$

Adding up these two inequalities side by side, one gets the limits between which the equation (40) is valid:

$$-1 < 2\,Y(t)\,dB(t) < 2 \,. \tag{45}$$

## 5.3. A numerical evaluation

We are going to check whether the CSL model predicts the results (**a**) and (**b**), by following the process in a proportional counter of cylindrical symmetry – figure 1 – containing a mixture of 95% Xe and 5% $CO_2$. The parameters of this counter, suitable for a number of PPC generations equal to 12, are detailed in the Appendix.

The tables I show the evolution of $/\boldsymbol{a}_1/^2$ along the detection process. As initial values are taken $/\boldsymbol{a}_1(t_0)/^2 = 0.4$, and $n_0 = 10$. $T_k$ denotes the time elapsed between the $k^{\text{th}}$ and the $(k+1)^{\text{th}}$ PPCs generation. It contains a couple of elementary intervals $dt$, and appears in the tables in units $dt$ – as said in section 4, $dt$ is chosen equal to $5 \times 10^{-13}\,\text{s}$. For each $T_k$, we replace $Y(t)$ by the RHS of (41). It is quite rough approximation, since not all the PPCs in a generation occur strictly simultaneously, though, what we seek out for the moment is a general image of the unfolding of the process.

The rest of the data in the $k^{\text{th}}$ column correspond to the end of the interval $T_k$. The symbol $\Sigma^{(k)}\text{sign}(\mathrm{d}B)$ denotes the sum of the signs of $\mathrm{d}B$ within $T_k$. Let's remind that in the section 4 we chose $\mathrm{d}B = \pm 2 \times 10^{-5}$.

For each elementary interval $\mathrm{d}t$, the sign of $\mathrm{d}B$ is picked arbitrarily. Thus, the $\mathrm{d}B$ signs in each $T_k$ are quite balanced, and so are along all the detection which lasts 201 elementary intervals: one can check that the sum of the quantities $\Sigma^{(k)}\text{sign}(\mathrm{d}B)$ over all the process, i.e. over all the values of $k$, is equal to 3.

The formulas (A11) and (A10) obtained in the Appendix A for calculating $T_k$, are repeated below:

$$T_1 = 12.84 \times 10^{-12}\,\text{s}, \quad \text{i)} \qquad T_{k+1} \approx 0.917\,T_k\,. \quad \text{ii)} \tag{46}$$

Beyond the column 11 the formula (40) cannot be used for arbitrary $\mathrm{d}B$ values in each $\mathrm{d}t$, because it would violate the inequality (45), i.e. would produce for $/a_l/^2$ values greater than 1, or negative. Though, I succeeded to avoid obtaining $/a_l/^2 > 1$ and $1 - /a_l/^2 < 0$ inside $T_{12}$, for some particular $\mathrm{d}B$ sets. As one can see, $/a_l/^2$ increased to 1. Also, calculating $\langle \hat{N} \rangle$ at the end of $T_{12}$ with the relation (36i), there resulted $\langle \hat{N} \rangle \approx Y$ (the exact result was 40959.59). Thus, the $\mathrm{d}B$ sequence used in the tables I, leads to the solution (**a**).[10]

## **Table I (part 1)**

| $k$ | 1 | 2 | 3 | 4 | 5 | 6 | 7 |
|---|---|---|---|---|---|---|---|
| $T_k$ | 26 | 23 | 22 | 20 | 18 | 17 | 15 |
| $\Sigma^{(k)}\text{sign}(\mathrm{d}B)$ | –4 | 1 | 0 | 2 | 0 | –1 | 3 |
| $Y_k$ | 20 | 40 | 80 | 160 | 320 | 640 | 1280 |
| $/a_l/^2$ | 0.4 | 0.4 | 0.4 | 0.4 | 0.4 | 0.39 | 0.42 |

## **Table I (part 2)**

| $k$ | 8 | 9 | 10 | 11 | 12 |
|---|---|---|---|---|---|
| $T_k$ | 14 | 13 | 12 | 11 | 10 |
| $\Sigma^{(k)}\text{sign}(\mathrm{d}B)$ | –2 | –1 | 2 | 3 | 0 |
| $Y_k$ | 2560 | 5120 | 10240 | 20480 | 40960 |
| $/a_l/^2$ | 0.4 | 0.28 | 0.3 | 0.98 | **0.99999** |

---

[10] The calculi for the tables were performed with the utility EXCEL, and the reader may request a copy of them by e-mail.

The tables II follow a different d$B$ sequence with arbitrary signs, which leads to the prediction (**b**).

## **Table II (part 1)**

| $k$ | 1 | 2 | 3 | 4 | 5 | 6 | 7 |
|---|---|---|---|---|---|---|---|
| $T_k$ | 26 | 23 | 22 | 20 | 18 | 17 | 15 |
| $\Sigma^{(k)}$sign(d$B$) | 0 | –1 | –2 | 0 | 4 | –1 | 1 |
| $Y_k$ | 20 | 40 | 80 | 160 | 320 | 640 | 1280 |
| $/\boldsymbol{a}_l/^2$ | 0.4 | 0.4 | 0.4 | 0.4 | 0.42 | 0.41 | 0.42 |

## **Table II (part 2)**

| $k$ | 8 | 9 | 10 | 11 | 12 |
|---|---|---|---|---|---|
| $T_k$ | 14 | 13 | 12 | 11 | 10 |
| $\Sigma^{(k)}$sign(d$B$) | 2 | 1 | –2 | –3 | 0 |
| $Y_k$ | 2560 | 5120 | 10240 | 20480 | 40960 |
| $/\boldsymbol{a}_l/^2$ | 0.45 | 0.47 | 0.19 | **3.4×10⁻⁵** | |

One can see that $/\boldsymbol{a}_l/^2$ practically vanishes, and so does $\langle \hat{N} \rangle$. The value obtained with the formula (36i) is 0.7, while by the theorem 1 it should be much less. Though, given the very approximate calculi done here it makes no sense to investigate the difference.

Here, the equation (40) could not be used beyond the column 11. This equation makes the transit from the quantum to the classical domain, so, the fact that its validity stops seems to tell that the system of particles became classical. Therefore, in continuation the system should be treated with the relation (41) which is purely classical. In support of that come the results in the tables I and II: before the equation (40) becomes invalid, $/\boldsymbol{a}_l/^2$ becomes practically equal to 1, respectively zero. Substituting in the formula (36i) the final result for $/\boldsymbol{a}_l/^2$ in these tables, one gets $\langle \hat{N} \rangle \approx Y$, respectively $\langle \hat{N} \rangle \approx 0$ as requires the theorem 1.

## **5.4. A difficulty**

Despite the fact that after $T_{11}$ the equation (40) doesn't hold anymore, not every d$B$ sequence leads to a physical result, (**a**) or (**b**).
Even a different arrangement of the sequence of signs of d$B$ within some interval $T_k$ in the tables I and II may lead to a totally different final result. See for instance in the tables III: the part 1 is identical with the part 1 of the tables I, so, it is not repeated here. In the part 2, the row $\Sigma^{(k)}\text{sign}(\mathrm{d}B)$ is the same as the corresponding row in the part 2 of the tables I. However, in the interval $T_{11}$ the sequence of signs of $\mathrm{d}B$ {1, 1, 1, −1, −1, 1, 1, −1, 1, −1, 1} was changed with the sequence {−1, 1, 1, −1, −1, 1, 1, −1, 1, 1, 1}. As one can see, the resulting $/\boldsymbol{a}_I/^2$ is completely different.

But, the essential fact is that after the process in the detector ended, there are no more PPC generations, though the result is neither $/\boldsymbol{a}_I/^2 \approx 1$ as in the tables I, nor $/\boldsymbol{a}_I/^2 \approx 0$ as in the tables II. What is the output of the detector in such a trial? And what about the other two wave-packets, are they erased, or not?

Another constraint on the response of the detector is that, in repeated trials, the physical results **(a)** and **(b)**, should appear with the probabilities predicted by the initial wave-function, $/\boldsymbol{a}_I/^2$, respectively $1-/\boldsymbol{a}_I/^2$. So, it seems that the dB series have to be controlled by the initial wave-function. But, it is not clear how could that be possible, because the value of $/\boldsymbol{a}_I/^2$ changes in time, apparently there is no memory of $/\boldsymbol{a}_I/(t_0)\,|^2$.

### **Table III (part 2)**

| $k$ | 8 | 9 | 10 | 11 | 12 |
|---|---|---|---|---|---|
| $T_k$ | 14 | 13 | 12 | 11 | 10 |
| $\Sigma^{(k)}\text{sign}(\mathrm{d}B)$ | −2 | −1 | 2 | 3 | 0 |
| $Y_k$ | 2560 | 5120 | 10240 | 20480 | 40960 |
| $/\boldsymbol{a}_I/^2$ | 0.4 | 0.28 | 0.3 | 0.44 | 0.34 |

In connection with this problem, P. Pearle said in [12], that the probability of a solution of the normalized CSL equation is given by the norm of the non-normalized equation, not by the probability of the dB sequence which drove the evolution of the wave-function.

"Let dΩ is the probability measure in the space of Brownian functions, and $w_\Omega$ be the particular Brownian function responsible for the evolution of the particular wavefunction $\psi_\Omega$, with squared norm

$$N_\Omega{}^2(t) = \int \mathrm{d}x \, /\psi_\Omega(x,t)/^2 \text{ [11]}$$

. . . the probability that $\psi_\Omega$ lies in the ensemble[12] is $N_\Omega{}^2(t)$, not dΩ.

. . . One expects that since dw (or dB) is as likely to fluctuate positively as negatively, the randomly fluctuating term[13] will have only a modest effect for the majority of Brownian motions . . . . . Therefore, the norms of the wavefunctions evolving subject to these Brownian motions will decrease roughly exponentially . . .

---

[11] Equation (8) in [12].
[12] In the ensemble of possible solutions.
[13] i.e. the term with dw or dB in the CSL equation – explanation of the present author.

On the other hand, for that minority of Brownian motions for which dw(x,t), or a set of dB(z,t) . . . happens to be of one sign significantly more frequently than of the other sign, the norm of the associated wavefunctions will grow . . ." [14]

However, the equation (40) in the present text has shown that the fluctuating sign of d$B$ does not diminish the influence of the noise, since this influence is augmented by $Y$, which increases all the time.

S. Adler proposed as a solution to the above problem, a proof, [35], by which the CSL model predicts that the square variance of the measured operator, tends to regain its initial value if the measurement takes a very long time, and if the square of the measured operator commutes with the Hamiltonian or the Hamiltonian is negligible in comparison with the other terms in the SSE. These conditions are not fulfilled in the general case. In our case the Hamiltonian cannot be neglected, as it drives the creation of electron-ion pairs. It contains creation and annihilation operators, e.g. (34). Neither does it commute with the square of the measured operator, i.e. with $\hat{N}^2 = (\hat{a}_e^\dagger \, \hat{a}_{ion}^\dagger \, \hat{a}_e \, \hat{a}_{ion})^2$.

## 5.5. Brief discussion of a related work

An analysis, related with the one presented in this text was done in [19]. A photon wave-function was supposed to contain two wave-packets, one of them hits a detector, the other doesn't. The configuration of the experiment is presented in figure 7. The authors declared:

"our essential assumption is that a measurement is completed by the time a permanent record of the event is made, that can be read out at a later time.
Each part of the setup has a different time scale, namely the detection time $t_D$, the amplification time $t_A$, and the recording time $t_R$. We thus identify the measurement time $t_M$ with the sum of these time scales."

The authors tried first to admit that the collapse is completed when the detector produces an output. That is logical, but they found that for the value they assigned to a parameter $\lambda$ ($\gamma$ in the present text) the collapse required much more time. So, they admitted that the collapse ends in the battery, which is non-physical. A battery is not a particle detector, it does not meet the wave-function. All it does is to push current in a circuit.

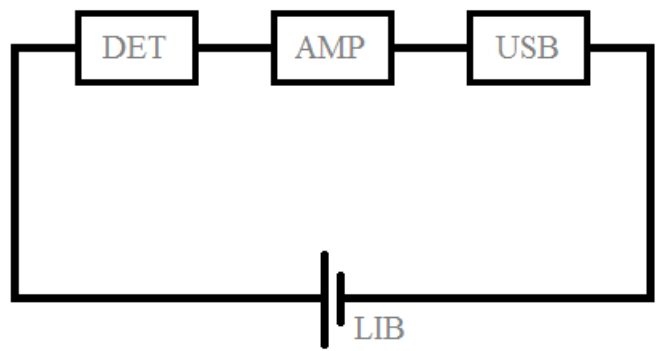

Figure 7. An alternative detection configuration.
The circuit contains a Lithium-ion battery (LIB), a detector (DET), an amplifier (AMP), and a flash drive that records the signal.

---

[14] The wave-functions Pearle discussed are non-normalized. To the difference, we work here with the solutions of (34), which are normalized.

Moreover, before the battery, neither the amplifier nor the display device (e.g. an oscilloscope) get as input a wave-function. One tunes the intensity of the input to these devices with a rheostat, not with a beam-splitter.

It is not clear why the authors didn't try to assign to $\lambda$ a greater value, given that in [34] they showed that the value of this parameter depends on experiment – is not a universal constant.

There are additional problems with their calculi,

Another problem: the amplification factor in a proportional ionization detector is not $10^{18}$ particles as the authors of [19] found with the battery explanation, but $10^3$ to $10^4$. It is sufficient for producing a classical signal, and since it is small, it may be amplified with highly sensitive classical amplifiers.

## 6. Conclusions

The CSL model for the measurement process of quantum systems has the advantage over different interpretations of the QM, that it does not propose changes in the QM formalism as long as the studied object consists only in one or a few microscopic components. In this case the SSE reduces to the Schrödinger equation, as the additional terms are negligibly small. However, as the number of components of the system increases, these terms increase in magnitude becoming more significant than the Hamiltonian. In this way is simulated the passing from the microscopic to the macroscopic objects.

The CSL model is not an *explanation* of the process of the wave-function collapse, as long as the nature of the noise incorporated in the SSE is not known. What it is clear, as explained in section 1, is that no classical field is the source of this noise. Thus, for the moment, this model represents a *tool* for investigating the collapse. In studying the noise, the test of the model on an experiment with a OPWF and a detector on only one of the wave-packets, is only a beginning, an incomplete (or, non-maximal) test, in the sense defined in [36]. It already left open a problem, described in the subsection 5.4. The model has to be tested on entanglements and on relativistic experiments. In this respect, the work of Bedingham [17] is not enough.

In this context I have to express criticism on a trend that seems to predominate at present: big efforts are invested in fitting Gaussian noises, non-white noises, for finding the values of the parameter $\gamma$, while beyond the corner await the entanglements and the relativity. These may impose requirements that the CSL might satisfy only by making hard compromises.

## Appendix

For each primary electron are liberated in the avalanche region a number $M$ of secondary electrons, number known under the name 'multiplication factor'

$$M = 2^{|V_a - V_1|/\Delta V}, \tag{A1}$$

where $V_1$ is the potential on the surface of the avalanche region, $V_a$ is the potential on the surface of the anode, and $\Delta V$ the potential fall along a path of an electron between two PPCs.

W. Diethorn calculated the potential fall $/V_a - V_1/$, and obtained – section III.B part 3 in [28],

$$/V_a - V_l/ = \frac{V_0}{\ln(b/a)} \ln\left[\frac{V_0}{a \, p \, K \ln(b/a)}\right], \tag{A2}$$

where $V_0$ is the potential difference between the anode and cathode, $a$ is the radius of the wire, $b$ the internal radius of the cathode, $p$ the pressure, and $K$ is the minimal field intensity per unit of pressure, at which a secondary ionization can occur.[15] For the mixture of gases mentioned in the subsection 5.3, the table 6.1 in chapter 6 section III.A of [26] indicates $\Delta V = 31.4\,\text{V}$ and $K = 36600\,\text{V/cm}\cdot\text{atm}$, so that with $a = 0.008\,\text{cm}$, $b = 1\,\text{cm}$, $V_0 = 1750\,\text{V}$, and $p = 0.4\,\text{atm}$, one obtains $/V_a - V_l/ \approx 400.5\,\text{V}$. Using this in (A1) there results, under the assumption (*) in section 5.1, that the number of PPC generations in the avalanche is 12. (In fact, $/V_a - V_l//\Delta V \approx 12.75$ i.e. the last PPC generation does not take place on the surface of the anode.)

With the numerical data mentioned above there results a multiplication factor $M = 4096$.

In the cylindrical geometry the relation between the potential at two points, at distances $r_1$, respectively $r_2$, from the central axis, is

$$\ln(r_2 / r_1) = -\frac{V_2 - V_1}{V_0} \ln(b / a). \tag{A3}$$

Setting $r_2 = b$, $r_1$ equal to the radius of the avalanche region, and $V_1 = 1750 - 400.5 \approx 1349.5\,\text{V}$, one gets $r_1 = 0.024\,\text{cm}$. From this radius inwards, we can make iteratively a rough evaluation of the radius of each PPCs region.

$$r_{k+1} / r_k = \exp\left[\frac{-\Delta V}{V_0} \ln(b / a)\right] = \exp\left[\frac{-31.4}{1750} \times 4.8283\right] \approx 0.917. \tag{A4}$$

The duration of the interval of flight between two PPCs can be calculated with the equality $\mathrm{d}r = v\,\mathrm{d}t$. As we will see below, the energies at which an electron is accelerated in the avalanche region render its wavelength so small that the movement of the wave-packet can be calculated with the classical kinematics. So, for the velocity $v(r)$ in the region $[r_k, r_{k+1}]$ we can use the formula

$$v(r) = \sqrt{2(V_r - V_{r_k})\,e / m_0}, \tag{A5}$$

$e$ being the elementary charge, $m_0$ the electron mass, and $r < r_k$.

With the help of (A3) we can translate the difference of potentials in (A5) into ratios of radiuses

$$v(r) = \sqrt{-2\,\frac{V_0\,e}{\ln(b / a)\,m_0} \ln(r / r_k)} \approx 1.077 \times 10^9 \sqrt{-\ln(r / r_k)}, \tag{A6}$$

the result being in units of cm/s. Thus, we get the equation

---

[15] The notation of the parameters here differs from that in [28], it is the same as in [26].

$$\mathrm{d}r / \sqrt{-\ln(r / r_k)} = 1.077 \times 10^9 \, \mathrm{d}t \,. \tag{A7}$$

This equation is difficult to integrate, so, let's try to simplify it. From (A4) one can see that r and $r_k$ are very close, therefore we can develop $\ln(r / r_k)$ in Taylor series of the quantity $r_k - r = \varepsilon$. Retaining only the first order of smallness, $-\ln[(r_k - \varepsilon) / r_k] \approx \varepsilon / r_k$. Thus, we obtain the differential equation

$$1.077 \times 10^9 \, \mathrm{d}t = \frac{-\mathrm{d}x}{\sqrt{1 - x}} r_k \,, \quad \text{i)} \qquad x = r / r_k \,. \quad \text{ii)} \tag{A8}$$

where the radius is in cm, and the time in seconds.

Integrating the RHS of (A8i) from $r_k$ to $r_{k+1}$, and the LHS from the time the electron reaches to radius $r_k$ to the time it reaches $r_{k+1}$, interval we denote by $T_k$, there results

$$1.077 \times 10^9 \, T_k = 2 \, r_k \sqrt{1 - r_{k+1} / r_k} \,. \tag{A9}$$

From this equation and from (A4) it's obvious that the following recursive relation emerges

$$T_{k+1} / T_k = r_{k+1} / r_k \approx 0.917 \,. \tag{A10}$$

However, we have first to calculate $T_1$. Setting in (A9) $k = 1$, using $r_1 = 0.024$ found above, one obtains

$$T_1 = 12.84 \times 10^{-12} \, \mathrm{s} \,. \tag{A11}$$

---